\documentclass[structabstract]{aa}  
\usepackage{graphicx}
\usepackage{txfonts}
\usepackage{wasysym}
\usepackage{ulem}
\begin{document}
   \title{The June 2012 transit of Venus.\\ Framework for interpretation
   of observations}

%   \subtitle{I. Overviewing the $\kappa$-mechanism}

   \author{}
   \author{A. Garc\'ia Mu\~noz
          \inst{1}%\fnmsep\inst{2}
          \and
          F. P. Mills\inst{2}%\fnmsep\thanks{Just to show the usage
          %of the elements in the author field}
          }

   \institute{Grupo de Ciencias Planetarias, Dpto. de F\'isica Aplicada I, 
    Escuela T\'ecnica Superior de Ingenier\'ia, Universidad del Pa\'is Vasco, 
    Alameda de Urquijo s/n, 48013 Bilbao, Spain\\   
    \email{tonhingm@gmail.com}
    \and
    Research School of Physics and
Engineering and Fenner School of Environment and Society, Australian
National University, Canberra, ACT 0200, Australia\\
    \email{frank.mills@anu.edu.au}
%             %\thanks{The university of heaven temporarily does not
%             %        accept e-mails}
    }

%   \date{Received September 15, 1996; accepted March 16, 1997}

% \abstract{}{}{}{}{} 
% 5 {} token are mandatory
 
  \abstract 
  {Ground-based observers have on 5--6th June 2012
  the last opportunity of 
  the century to watch the passage of Venus across the solar disk from Earth.
  Venus transits have traditionally provided unique insight into 
  the Venus atmosphere through the refraction halo that appears at the planet's
  outer terminator near ingress/egress. 
  Much more recently, Venus transits have attracted renewed interest because
  the technique of transits is being succesfully applied to the characterization of
  extrasolar planet atmospheres.
  }
  {The current work investigates  theoretically 
  the interaction of sunlight and the Venus
  atmosphere through the full range of transit phases, as observed
  from Earth and from a remote distance.
  Our model predictions quantify the relevant atmospheric phenomena, thereby
  assisting the observers of the event in the intepretation of measurements 
  and the extrapolation to the exoplanet case.
  }
  {Our approach relies on the numerical integration of the radiative transfer 
  equation, and includes refraction, 
  multiple scattering, atmospheric
  extinction and solar limb darkening, as well as an up-to-date description of 
  the Venus atmosphere. 
  }
  {We produce synthetic images of the planet's terminator during ingress/egress
  that demonstrate the evolving shape, brightness and chromaticity 
  of the halo. Our simulations reveal the impact of micrometer-sized
  aerosols borne in the upper haze layer of the atmosphere on the halo's 
  appearance. 
  Guidelines are offered for the investigation of the planet's upper haze
  from vertically-unresolved photometric measurements.
  In this respect, the comparison with measurements from the 2004 transit 
  appears encouraging.
  We also show integrated lightcurves of the Venus-Sun system at various phases 
  during transit and calculate the respective Venus-Sun integrated 
  transmission spectra.
  The comparison of the model predictions to those for a Venus-like planet
  free of haze and clouds (and therefore a closer terrestrial analogue) 
  complements the discussion and sets the conclusions
  into a broader perspective. 
  }
   {}


   \titlerunning{The June 2012 transit of Venus}
   \maketitle
%
%________________________________________________________________

\section{Introduction}

This year, from about 22:10UT June 5 to 04:50UT June 6,
observers across the Earth's illuminated hemisphere have the opportunity to
watch Venus passing in front of the solar disk. 
The transits of Venus visible from  Earth are regularly spaced in intervals of 
8, 105.5, 8 and 121.5 years. %\footnote{F.
%Espenak, NASA/GSFC, http://eclipse.gsfc.nasa.gov/eclipse.html} 
After the June 2004 transit, this year's event 
is the last one before 2117. 

Venus transits occupy a notable place in the history of astronomy
and planetary sciences.
Following Halley's suggestion that Venus transits might provide a measure of the
terrestrial parallax, the events 
of the 18th and 19th centuries helped estimate the Earth-Sun distance. 
Also, it is often quoted that the 1761 transit provided the first evidence for
the Venus atmosphere (Link, \cite{link1959}).
Lomonosov correctly attributed the halo that appears at the planet's outer 
terminator during ingress/egress to sunlight rays refracted towards the Earth 
on their passage through the Venus atmosphere.
The subsequent transits of 1769, 1874 and 1882 provided additional
opportunities for the investigation 
of the phenomenon and, in turn, of the Venus atmosphere. 
Since then, a number of theories have been issued
to interpret the halo and its connection with the Venus atmospheric structure.
A discussion of them and of other phenomena related to refraction in planetary
atmospheres is presented by Link (\cite{link1969}).

An additional reason for pursuing planetary transits %in the Solar System 
has emerged recently. % , driving enhanced interest in the 2004 and 2012 Venus transits.
The discovery of planets in orbit around stars other than our Sun 
has opened up a new and rapidly-growing field in astrophysics 
(Mayor \& Queloz, \cite{mayorqueloz1995}). 
Some of those planets happen to 
periodically intercept the Earth-to-star line of sight, thereby causing a
dimming in the apparent stellar brightness. 
%To date, the so-called technique of transits has successfully been applied 
%to the detection or confirmation of more than 200 extrasolar
%planets.\footnote{J. Schneider, CNRS/LUTH -- Paris Observatory, 
%The Extrasolar Planets Encyclopaedia, www.exoplanet.eu} 
%%from ground- and space-based observatories.\footnote{} 
 In combination with some form of spectral discrimination, 
the so-called technique of transits may provide valuable insight into the 
composition and state of the planets' atmospheres (\textit{e.g.} Seager \&
Deming, \cite{seagerdeming2010}, and refs. therein). 
In our Solar System, only Mercury and Venus are observable from Earth while 
transiting the Sun. Since Mercury's atmosphere is tenuous, 
the Venus transits of 2004 and 2012 represent unique 
occasions for testing the technique of transits on an Earth-sized planet hosting
a dense atmosphere, 
a major goal in the near-future exploration of extrasolar planets.

%\textbf{Why is it important to investigate Earth-size exoplanets?? Life as we 
%know it can only exist on such planets??} 

The interest raised by the 2004 and 2012 Venus transits
%and the various associated physical phenomena 
is apparent %in the number of recent publications
(Ambastha et al., \cite{ambasthaetal2006}; 
Ehrenreich et al., \cite{ehrenreichetal2012}; 
Hedelt et al., \cite{hedeltetal2011}; 
Kopp et al., \cite{koppetal2005}; 
Pasachoff et al., \cite{pasachoffetal2011}; 
Schneider et al., \cite{schneideretal2004, schneideretal2006}; 
Tanga et al., \cite{tangaetal2012}). 
An ambitious international plan of observations from 
both Earth-based and space-borne
observatories is underway for the 2012 event (Pasachoff, \cite{pasachoff2012}). 
It thus seems appropriate to review the theory of planetary transits
in its specific application to the Venus transit, taking advantage of recent
progress in the characterization of the Venus atmosphere and of 
modern computational capacity.

This work addresses the interaction of sunlight and the Venus atmosphere during
transit. 
%phenomena that produce the
%signature of the Venus atmosphere on the emitted solar light during transit.
%Our approach deals with all phases of the event (out of transit, ingress/egress 
%and in transit) in a unified manner.
% by direct solution of the Radiative Transfer
%Equation (RTE) at the observer's site. % for the direct and diffuse sunlight components. 
Descriptions of the formulation and the prescribed model parameters
are given in {\S}\ref{theory_sec}. 
Section {\S}\ref{halo_sec} investigates the shape, 
brightness and chromaticity of
the halo. 
Section {\S}\ref{transit_sec} establishes the connection between the halo
 and the in-transit signature of the Venus atmosphere, and
presents the lightcurves of the Venus-Sun system as observed from 
Earth and remotely.
Section {\S}\ref{conclusions_sec} summarizes the main predictions. 
Ultimately, it is hoped that our model predictions can serve as guidelines 
 in the interpretation of the observations to come. 
 
\section{\label{theory_sec} Refractive theory of the Venus transit}
 
Lunar eclipses and planetary transits are 
conceptually similar phenomena that admit a 
common formal treatment (Link, \cite{link1969}). 
Garc\'ia Mu\~noz \& Pall\'e (\cite{garciamunozpalle2011}) 
revisited the theory of lunar eclipses. Their formulation was subsequently 
used in the
interpretation of a partial lunar eclipse
(Garc\'ia Mu\~noz et al., \cite{garciamunozetal2011}), and 
to investigate the impact of refraction 
on the in-transit signature of Earth-like extrasolar planets
(Garc\'ia Mu\~noz et al., \cite{garciamunozetal2012}).
We now apply the formulation to the Venus transit. 
Since the refractive theory of planetary transits has thus far received marginal
attention, we expend some effort describing the model approach.

\subsection{\label{rte_sec}The Radiative Transfer Equation (RTE)}

The steady-state RTE for a refractive medium 
in the geometrical optics limit has been presented by, 
\textit{e.g.}, Zheleznyakov (\cite{zheleznyakov1967}) and 
\'Enom\'e (\cite{enome1969}): 
\begin{equation} 
\mathbf{s}\cdot \nabla \left( \frac{L\mathbf{(x,s)}}{n^2\mathbf{(x)}} \right)
=-\gamma(\mathbf{x}) 
\frac{L\mathbf{(x,s)}}{n^2\mathbf{(x)}} +
\frac{J\mathbf{(x,s)}}{n^2\mathbf{(x)}}.
\label{integrodiff_eq} 
\end{equation}
Here, 
$L\mathbf{(x,s)}$ is the radiance 
%(in, for instance, units of photons s$^{-1}$cm$^{-2}$sr$^{-1}$(cm$^{-1}$)$^{-1}$)
at $\mathbf{x}$ in direction ${\bf{s}}$, 
$\gamma(\mathbf{x})$ and $n(\mathbf{x})$ stand for the 
extinction coefficient and index of refraction of the medium, respectively,
and $J\mathbf{(x,s)}$ is the source term, which may include separate terms
for scattered radiation and emission from within the medium. 
$L\mathbf{(x,s)}$, $\gamma(\mathbf{x})$, $n(\mathbf{x})$ and other
associated magnitudes are wavelength dependent, which should be explicitly
stated in the notation.
For simplicity, however, that characteristic is omitted throughout the text. 
The boundary conditions for the direct and diffuse components of 
Eq. (\ref{integrodiff_eq}) are introduced in 
{\S}\ref{direct_sec} and \ref{diffuse_sec}, respectively.

For $n(\mathbf{x})$ constant,  Eq. (\ref{integrodiff_eq}) turns into 
the well-known RTE for a non-refractive medium.
In that case, the photons follow straightline trajectories between scattering
events.
For a non-absorbing, non-emitting, non-scattering medium 
($\gamma(\mathbf{x})$$\equiv$$J(\mathbf{x})$$\equiv$0)  
with possibly spatially-dependent $n(\mathbf{x})$, 
Eq. (\ref{integrodiff_eq}) reduces to
the conservation of ${L\mathbf{(x,s)}}/{n^2\mathbf{(x)}}$ on the
refraction-bent ray trajectories. 
That magnitude is sometimes referred to as the Clausius invariant,
and generalizes 
the conservation of radiance in media of constant index of refraction.
%(Fumeron \& Asllanaj, \cite{fumeronasllanaj2009}).

As in radiometry of non-refractive media, 
the irradiance, %(in units of photons s$^{-1}$cm$^{-2}$(cm$^{-1}$)$^{-1}$)  
or net flux of light that crosses an elementary surface of normal vector 
${\bf{n_{x_O}}}$ 
(not to be mistaken for the index of refraction at $\mathbf{x_O}$, 
$n(\mathbf{x_O})$) 
from directions $\mathbf{s_O}$,
at the observer's location $\mathbf{x_O}$ is obtained by integrating 
the directional radiance ${{L}}({\bf{x_O}}, {\bf{s_O}}) $
over solid angle: % $d{\Omega(\bf{s_O})}$:
\begin{equation}
F({\bf{x_O}})=\int_{\partial\Omega=\partial\Omega_{\venus}\cup\partial\Omega_{\odot}}{ 
               {{L}}({\bf{x_O}}, {\bf{s_O}}) 
               {{\bf{s_O}}\cdot{\bf{n_{x_O}}}}
	       d{\Omega(\bf{s_O})}.
} 
\label{irradiance1_eq}
\end{equation}
In the general application of Eq. (\ref{irradiance1_eq}) to the Venus transit, 
the integration domain %, 
% $\partial\Omega$=$\partial\Omega_{\venus}$$\cup$$\partial\Omega_{\odot}$,  
must contain the solid angles subtended from $\mathbf{x_O}$ by planet and 
atmosphere together, % up to a specified altitude, 
$\partial\Omega_{\venus}$, and the Sun, $\partial\Omega_{\odot}$. 

For the numerical evaluation of Eq. (\ref{irradiance1_eq}), it is convenient to
work with auxiliary variables. % that can be readily discretized.
Concentrating for now on the integral over $\partial\Omega_{\venus}$, 
introducing: 
$$\mu={\bf{s_O}}\cdot{\bf{n_{x_O}}},$$
and the azimuthal angle $\phi$, using: 
$$d{\Omega(\bf{s_O})}=d\mu d\phi,$$ 
and the normalized variables:
$$
\xi=(\mu^2-\mu^2_{\rm{M}})/(\mu^2_{\rm{m}}-\mu^2_{\rm{M}} )\;\; \rm{and}
\;\;\eta={\phi}/{\pi}, 
$$ 
Eq. (\ref{irradiance1_eq}) can be rewritten as: 
\begin{equation} 
F({\bf{x_O}})=\pi (\mu^2_m-\mu_M^2) \int_0^1 \int_0^1 L(\xi,\eta)
d\xi d\eta, 
\label{irradiance4_eq}
\end{equation} 
where: 
$$\mu^2_{\rm{M}}=1-(r_{\rm{TOA}}/d_{\rm{\bf{O}}})^2,$$ 
$$\mu^2_{\rm{m}}=1-(R_{\venus}/d_{\rm{\bf{O}}})^2,$$
$$r_{\rm{TOA}}=R_{\venus}+h_{\rm{TOA}},$$
$R_{\venus}$ is the Venus mean radius, 
$h_{\rm{TOA}}$ is the top of the atmosphere altitude, 
and $d_{\rm{\bf{O}}}$ is the observer-to-Venus centre distance.
We may also define the impact radius $r_b$ for each $\mu^2$ 
through: 
$$\mu^2=1-(r_b/d_{\rm{\bf{O}}})^2.$$
Here, $r_b$ has the meaning of the
distance of closest approach to the Venus centre for straightline rays traced
sunwards from $\mathbf{x_O}$. 
Essentially, Eq. (\ref{irradiance4_eq}) maps the original integral
onto a plane of incident directions 
$\mathbf{s_O}$=\{$\xi, \eta$\} sampling the field of view 
from $\mathbf{x_O}$. 
Figure 1 in Garc\'ia Mu\~noz \& Pall\'e 
(\cite{garciamunozpalle2011}) sketches the meaning of some of the variables
introduced above.

\subsection{\label{direct_sec}The direct component of sunlight}

\begin{figure*}
   \centering
   \includegraphics[width=9cm]{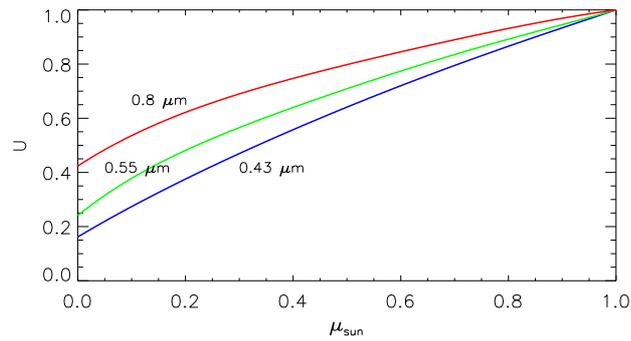}
      \caption{Limb darkening function $U$($\mu_{\odot}$) at selected
      wavelengths. For a color version of the figure, see the electronic
      journal.}
         \label{U_fig}
% /VENUSTRANSIT/LIMBDARKENING/UU.pro
   \end{figure*}

The usual treatment of the RTE separates
the direct and diffuse components. 
Neglecting the emission from within the medium and 
taking ${J\mathbf{(x,s)}}$$\equiv$0, one obtains
for the direct radiance: 
\begin{equation}
\frac{L_D\mathbf{(x_O,s_O)}}{n^2\mathbf{(x_O)}}=
\frac{L_D\mathbf{(x_\odot,s_\odot)}}{n^2\mathbf{(x_\odot)}} 
 \exp{[-\int_{\Gamma} \gamma{(s)} ds]},
\label{Ln2_eq}
\end{equation}
for ray trajectories $\Gamma$ connecting $\mathbf{x_O}$ and the solar disk. Here, 
$s$ is the path length on the ray trajectory and 
${\bf{x_\odot}}$ and ${\bf{s_\odot}}$ are the location and direction of 
departure of the sunlight ray on the solar disk. 
The squared refraction indices at the observer's site and at the Sun's
photosphere, ${n^2\mathbf{(x_O)}}$ and ${n^2\mathbf{(x_\odot)}}$ respectively,
can be taken as 
$\approx$1 in Eq. (\ref{Ln2_eq})
at visible and near-infrared wavelengths. 
$L_D\mathbf{(x_\odot,s_\odot)}$ is the radiance emitted by the Sun and enters 
the formulation as a boundary condition. 
We assume that it depends on the emission angle, $\cos^{-1}(\mu_\odot)$, 
through a polynomial of the form $U$= $\sum_{i=0}^5 u_i \mu^i_{\odot}$, 
and that the wavelength-dependent coefficients $u_i$ account for the
chromatic effects in the solar output.
A good representation of the solar output near the Sun's edge, 
$\mu_{\odot}$$\rightarrow$0, is important for the simulation of the
ingress/egress phases of the transit.
The adopted wavelength-dependent limb-darkening coefficients  $u_i$ are from 
Pierce \& Slaughter (\cite{pierceslaughter1977}) and Pierce et al. 
(\cite{pierceetal1977}).
Generally, the coefficients are dependent on the star's metallicity, a fact that must
properly be accounted for in the analysis of extrasolar planet lightcurves 
(Claret, \cite{claret2000}; Sing, \cite{sing2010}). 
Thus, $L_D\mathbf{(x_\odot,s_\odot)}$=$B_{\odot} U$, where 
$B_{\odot}$ is the stellar radiance at the Sun's center, $\mu_{\odot}$=$1$. 
(Note also that $\sum_{i=0}^5 u_i$=1.) 
Figure (\ref{U_fig}) shows the implemented 
$U$($\mu_{\odot}$) at selected wavelengths.

Considering the above, to a good approximation:
\begin{equation}
L_D\mathbf{(x_O,s_O)}=
B_{\odot} U \exp{[-\int_{\Gamma} \gamma{(s)} ds]},
\label{Ln3_eq}
\end{equation}
which is the Beer-Lambert law of extinction in a
non-refractive medium. 
Using Eq. (\ref{Ln3_eq}) 
does not mean that refraction plays a minor role in the formulation.
Refraction enters $L_D\mathbf{(x_O,s_O)}$ at the time
of devising the ray trajectory $\Gamma$. 
For that purpose,  %and for the determination of the ray trajectories, 
we use the tracing scheme based on path length by 
van der Werf (\cite{vanderwerf2008}) and trace each ray trajectory from 
${\bf{x_O}}$ sunwards after specifying the Venus atmospheric profile for the 
wavelength-dependent refraction index.

For the evaluation of Eq. (\ref{irradiance4_eq}) 
over %the solid angle subtended by the planet and atmosphere, 
$\partial\Omega_{\venus}$, the $\xi$--$\eta$ field of view at $\mathbf{x_O}$ 
is subdivided into
$N_{\xi}$$\times$$N_{\eta}$ (=200$\times$360) evenly-sized bins, 
each of them associated with an
incident direction $\mathbf{s_O}$. % at $\mathbf{x_O}$. 
%We typically take $N_{\xi}$$\times$$N_{\eta}$=1000$\times$360.  
For each direction $\mathbf{s_O}$=\{$\xi, \eta$\} a ray trajectory $\Gamma$ is traced, 
and the corresponding optical opacity going into Eq. (\ref{Ln3_eq})
is calculated. 
The rays failing to connect $\mathbf{x_O}$ with the solar disk have  
$L_D\mathbf{(x_O,s_O)}$$\equiv$0. 
In the halo, during ingress/egress, 
$L_D\mathbf{(x_O,s_O)}$$\equiv$0 occurs for rays crossing 
nearly unrefracted through the upper layers of the atmosphere, 
but also for rays passing through the lowermost altitudes and
whose trajectories become deflected away from the Sun. 
During transit, the trajectories failing to connect 
$\mathbf{x_O}$ with the solar disk impose a ring of 
altitudes at the planet's terminator 
that cannot be probed from a remote distance. The latter is 
discussed in detail by Garc\'ia Mu\~noz et al. (\cite{garciamunozetal2012}) and 
will be noted again for the specific case of Venus in 
\S\ref{geometrical_sec} and \ref{diskintlightcurve_sec}. 
By going through the entire $N_{\xi}$$\times$$N_{\eta}$ matrix of 
directions ${\bf{s_O}}$, 
the image of the solar disk at ${\bf{x_O}}$ 
emerging from the planet's terminator is constructed.

We could adopt the same strategy to evaluate Eq. 
(\ref{irradiance4_eq}) over the rest of $\partial\Omega$ not contained in 
$\partial\Omega_{\venus}$.
This is, however, rather inefficient given the large 
ratio of Sun/Venus radii, $R_{\odot}/R_{\venus}$.
Instead, we integrate analytically the irradiance
over the unobstructed view of the star, 
$F_{\odot}$=$\pi$$(R_{\odot}/l_{\odot})^2$$\sum_{i=0}^5 2 u_i / (i+2)$, 
where $l_{\odot}$ is the observer-to-Sun centre distance, and
subtract the part corresponding to $\partial\Omega_{\venus}$ calculated by Monte
Carlo integration.
 
%However \textit{picturesque} (Link, \cite{link1962}), 
Taking the observer's location $\mathbf{x_O}$ for carrying out 
the integration of Eq. (\ref{irradiance4_eq}) 
for the direct sunlight component is in line with the early formulations of the 
photometric theory of lunar eclipses
(Link, \cite{link1962}, \cite{link1969}). Link (\cite{link1969}), 
in Section 1.2.21 of that monograph, comments on the 
equivalence of that approach with other formulations that alternatively
carry out the integration at the solar plane.
Link (\cite{link1969}) also refers to the so-called 
attenuation by refraction that occurs 
by the compression of the solar disk image at the planet's
terminator during ingress/egress, 
and notes that its effect is implicitly accounted for in the
projection of $d\Omega(\mathbf{s_O})$ on the solar disk.
Attenuation by refraction occurs
always that sunlight rays traverse a planet's 
stratified atmosphere (Hays \& Roble, \cite{haysroble1968}), and merely
represents the change in cross section undergone by a pencil of sunlight rays 
upon crossing the stratified medium.

\begin{figure*}
   \centering
   \includegraphics[width=9cm]{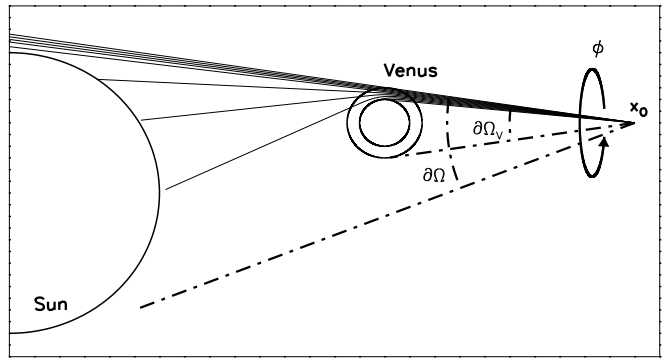}
      \caption{ Sketch of the Venus-Sun system at an instant 
      near internal contact. 
      The rays traced from ${\bf{{x_O}}}$ fan out as they traverse the 
      Venus stratified atmosphere. 
      }
         \label{sketch_fig}
% /VENUSTRANSIT/SKETCH
% /VENUSTRANSIT/XtraRT_FIG1
   \end{figure*}

\begin{figure*}
   \centering
   \includegraphics[width=8cm]{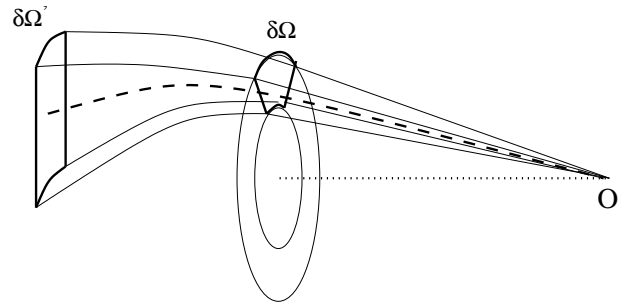}
      \caption{
      The sketch demonstrates how a differential element of solid angle 
      $d\Omega$ fans out when traced sunwards from the observer's site. 
      The dashed line represents the trajectory passsing through 
      the middle-point of the solid angle element. In a transparent atmosphere,
      the ratio of radiance and the squared refraction index is conserved on the
      ray trajectory, and the corresponding irradiance for the differential
      solid angle element is simply 
      ${{L}}({\bf{x_O}}, {\bf{s_O}}) {{\bf{s_O}}\cdot{\bf{n_{x_O}}}}
	       d{\Omega(\bf{s_O})}$$\sim$
      ${{L}}({\bf{x_{\odot}}}, {\bf{s_{\odot}}}) d{\Omega(\bf{s_O})}$. 
      For the calculation of Eq. (\ref{irradiance4_eq}), a mathematical
      transformation is applied to rewrite the integral in the 
      the auxiliary variables $\xi$ and $\eta$.
      }
         \label{newsketch_fig}
   \end{figure*}

Figure (\ref{sketch_fig}) offers some visual insight into this point. 
The rays in the sketch are evenly spaced in angle 
at their departure from $\mathbf{x_O}$ towards the Sun.
The rays gradually separate as they traverse the atmosphere 
due to differential refraction between altitude layers. 
Assuming that the atmosphere is transparent and solar limb darkening 
is omitted, Eq.
(\ref{irradiance4_eq}) would simply lead to a measure of the halo's angular
size perceived at $\mathbf{x_O}$.
% extent in the $\xi$--$\eta$ plane. 
Since in such conditions the radiance is very approximately conserved on ray
trajectories, 
the contribution of a solar disk surface element %is not dictated by its unrefracted size but it 
is ultimately limited by the size of its refracted image at the planet's 
terminator. 
Sidis \& Sari (\cite{sidissari2010}) set out from this conclusion to evaluate
the halo's overall brightness for transiting extrasolar giant planets
with transparent atmospheres. 
The factor by which the pencil's cross section 
is compressed depends on both the Sun-Venus and Venus-observer
distances as well as on the atmosphere's refractive properties.
Figure (\ref{newsketch_fig}) sketches the integration of the direct
radiance over a differential element of solid angle at the observer's site.

\subsection{\label{diffuse_sec}The diffuse component of sunlight}

A fraction of the sunlight photons incident on Venus is
redirected in the forward direction 
after one or more scattering collisions with the medium.
To estimate the contribution of diffuse sunlight, 
we must solve the RTE including:
\begin{equation} 
J\mathbf{(x,s)}=
\beta(\mathbf{x}) 
 \int_{\partial \Omega_{\venus}}  p\mathbf{(x,s,s')}L\mathbf{(x,s')}
 d\Omega(\mathbf{s'})
\label{J_eq} 
\end{equation}
as a source term of the equation, 
with %$L_d\mathbf{(x,s)}$ denotes specifically the diffuse radiance, 
$\beta(\textbf{x})$ being the medium's scattering coefficient
and $p\mathbf{(x,s,s')}$ the normalized phase function
for changes $\mathbf{s'}$$\rightarrow$$\mathbf{s}$ in the direction of light
propagation. % normalized by $\int$${p d\Omega}$=1. 
The given $J\mathbf{(x,s)}$ is identical to the form of the scattering term in 
non-refractive media
(Ben-Abdallah et al., \cite{benabdallahetal2001}; Kowalski \& Saumon,
\cite{kowalskisaumon2004}; Fumeron et al., \cite{fumeronetal2005}). 
The integration over the part of $\partial \Omega$ not included in 
$\partial \Omega_{\venus}$ is unnecessary if the top of the atmosphere is placed
high enough.

Working in a refractive medium presents two additional subtleties in the treatment of
the diffuse RTE.
First, the radiance is affected by the
reciprocal of $n^2(\mathbf{x})$,  
although 
we are allowed to take $n(\mathbf{x})$$\approx$1 at the relevant
wavelengths. 
Second, the photon trajectories in between scattering events follow 
refraction-bent trajectories. 
However, scattering itself alters the direction of light propagation by
angles likely to exceed the deflection caused by refraction. 
We may therefore also omit this aspect, 
and treat the diffuse RTE as in a non-refractive medium.

The practical solution to Eqs. (\ref{integrodiff_eq}) and (\ref{J_eq}) 
is carried out by Monte Carlo integration  (O'Brien, 
\cite{obrien1992}; Garc\'ia Mu\~noz \& Pall\'e, \cite{garciamunozpalle2011})
with the proper expression for $F_{\odot}$ at the Venus orbital distance
as a boundary condition. 
The scheme is nested in the simultaneous integration of 
Eq. (\ref{irradiance4_eq}) over the Venus disk.

\subsection{\label{venusatmosphere_sec} The reference Venus atmosphere}

Traditionally, observations of Venus during ingress/egress have provided valuable
information on the tilt of the planet's rotational axis and 
equator-to-pole variations in the atmospheric structure (Link, \cite{link1959}).
Thus, for the simulations we require 
a description of the atmospheric optical properties
on a global scale as well as the latitudinal dependence for some of them.

For the temperature, 
we adopted the Seiff et al. 
(\cite{seiffetal1985}) profile for a latitude of 45$^{\circ}$ up to 100 km 
and the Hedin et al. (\cite{hedinetal1983}) noon profile there upwards. 
The atmosphere was assumed to be composed of CO$_2$--N$_2$
in a 96.5--3.5\% proportion, and the background 
density was obtained by integrating the hydrostatic balance equation.

At the altitudes at which the atmosphere becomes optically thick, the opacity
is largely dictated by aerosols rather than by the atmospheric gas. Thus, 
latitudinal variations in the gas profile are expected to be of less importance 
than the respective variations in the aerosol loading. Similarly, the errors
introduced in the refracted angles from neglecting the
latitudinal variation in the density profiles are minor.

The gas refractivity is needed to trace the ray 
trajectories and to calculate the Rayleigh scattering cross sections. 
We adopted for CO$_2$ the fit between 0.48 and 1.82 $\mu$m by 
Old et al. (\cite{oldetal1971}), and for N$_2$ the fits 
between 0.2 and 2.1 $\mu$m by Bates (\cite{bates1984}).
The fits were truncated at 2 $\mu$m at the
longer wavelengths and extrapolated at the shorter wavelengths.

The Venus atmosphere is heavily loaded with aerosols that become 
optically thick in the nadir at about 70 km.
 By convention, aerosols above and below the optical thickness $\tau_{\rm{nadir}}$=1
are usually referred to as haze and cloud particles,
respectively (Esposito et al., \cite{espositoetal1983}). 
Limb opacities are enhanced by a geometrical factor (Barth, \cite{barth1969}) of
$\sim$$\sqrt{2\pi R_{\venus}/H_{\rm{aer}}}$$\sim$100, estimated with an aerosol  
scale height $H_{\rm{aer}}$$\sim$4 km, which means that 
we will hereon only consider the aerosol opacity due to haze.

\begin{figure*}
   \centering
   \includegraphics[width=9cm]{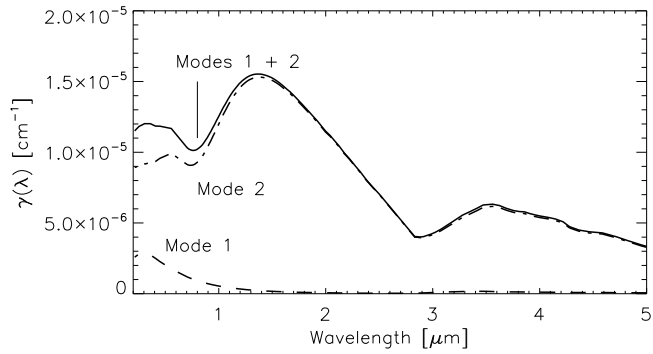}
      \caption{Haze extinction coefficient at 60 km for 
      cloud top at 70 km. At other altitudes, 
      $\gamma(\lambda; z)$=$\gamma(\lambda; 60\;\rm{km})$$\times$$p/p(60\;\rm{km})$.
      }
       \label{gamma60km_fig}
% /VENUSTRANSIT/HAZE/pinta_haze.pro
   \end{figure*}

\begin{figure*}
   \centering
   \includegraphics[width=9cm]{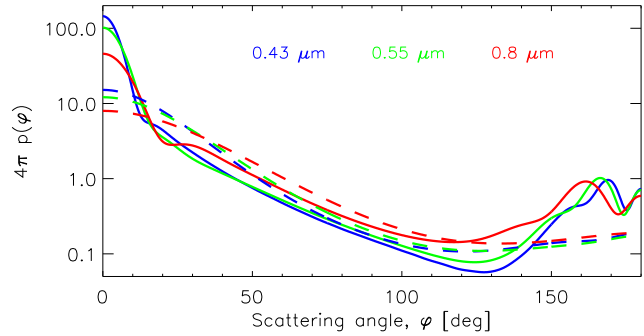}
      \caption{Scattering phase function for mode 1 (dashed) and mode 2 (solid)
      haze at selected wavelengths. A scattering angle of zero corresponds to
      the forward scattering direction. 
      For a color version of the figure, see the electronic
      journal.}
       \label{phF_fig}
% /VENUSTRANSIT/DIFFUSE/pinta_phf.pro
   \end{figure*}

To calculate the haze opacity we followed the prescriptions by 
Molaverdikhani et al. (\cite{molaverdikhanietal2012}), which are consistent with
those by Pollack et al. (\cite{pollacketal1980}). 
Thus, from 60 to 80 km, we adopted 
log-normal size distributions of aerosols that depend on
the particle's radius $r_a$ through:
$$n_a(r_a)= \frac{1}{\sqrt{2\pi} \ln{\sigma_g}} \frac{1}{r_a}
\exp{(-(\ln{r_a}-\ln{r_g})^2/2\ln^2{\sigma_g})}.$$  
Mode 1 haze is characterized by $r_g$=0.15
$\mu$m and $\sigma_g$=1.5 (effective radius and variance
of $r_{\rm{eff}}$=0.23
$\mu$m and $v_{\rm{eff}}$=0.18, respectively), and a number density profile  
that drops with the
background pressure as 900$\times$$p$/$p(60\;\rm{ km})$ cm$^{-3}$. 
For mode 2 haze, $r_g$=1.0
$\mu$m, $\sigma_g$=1.21 ($r_{\rm{eff}}$=1.09
$\mu$m and $v_{\rm{eff}}$=0.037), and the number density profile  
is 120$\times$$p$/$p(60\;\rm{km})$ cm$^{-3}$. 
For our simulations, we extended indefinitely upwards the $p$/$p(60\;\rm{ km})$ 
profile. Indeed, mode 1 particles are long known to exist 
above 80 km, and two works (Wilquet et al., \cite{wilquetetal2009}; 
de Kok et al., \cite{dekoketal2011})
recently identified 
sizeable amounts of mode 2 particles up to 90--95 km.
Both $r_{\rm{eff}}$ and $v_{\rm{eff}}$ are defined by Hansen \& Travis
(\cite{hansentravis1974}). 
Note, though, that the definition of $\sigma_g$ in
that work corresponds to $\ln{\sigma_g}$ here.
We assessed that the effect of extending the aerosol profile above 100 km
was negligible because the aerosol layer becomes effectively thin at such
altitudes.
The wavelength-dependent cross sections, albedos and scattering
phase functions for each mode are calculated from Mie
theory (Mishchenko et al., \cite{mishchenkoetal2002}). %\footnote{Also: 
%M. I.  Mishchenko, \\
%http:$//$www.giss.nasa.gov$/$staff$/$mmishchenko$/$t$\_$matrix.html.
%}
 The complex refractive indices for the aerosol particles 
(H$_2$SO$_4$/H$_2$O solutions at 84.5\% by weight, 
Molaverdikhani et al. (\cite{molaverdikhanietal2012})) were borrowed from 
Palmer \& Williams (\cite{palmerwilliams1975}). 
Figure (\ref{gamma60km_fig}) graphs the extinction profile at 60 km from 
0.2 to 5 $\mu$m, that appears to be dominated by mode 2 particles.
Figure (\ref{phF_fig}) shows the scattering phase function of each 
mode at specific wavelengths. 

The aerosol profile %described above
 based on Molaverdikhani et al. (\cite{molaverdikhanietal2012}) 
reaches $\tau_{\rm{nadir}}$=1 at 1.6 $\mu$m near 70 km, which may 
represent a globally-averaged cloud top altitude. 
A limb opacity $\tau_{\rm{limb}}$=1 
is reached about 4--5 scale heights higher and therefore likely above 85 km. 
This globally-averaged description is correct to first order, but the
appropriate simulation of the halo requires a more accurate characterization of
latitudinal variations in the cloud top level. 
Recent altimetry of the Venus clouds from Venus Express 
shows that $\tau_{\rm{nadir}}$=1 is reached at about 74 km 
near the equator 
but may drop to 64 km at the poles, with transition zones on both hemispheres 
at about 60$^{\circ}$ latitude (Ignatiev et al., \cite{ignatievetal2009}). 
To account for this latitudinal dependence, 
we shifted at each latitude the aerosol profiles upwards or downwards 
as needed to match the cloud top altitudes graphed in Fig. 8b of 
Ignatiev et al. (\cite{ignatievetal2009}).

\section{\label{halo_sec}The halo}

The halo (or aureole) is the refracted image of the Sun that forms at the planet's outer
terminator for small Sun-Venus angular separations, with possibly 
a minor contribution from diffuse sunlight. 
Its appearance is determined by 
the geometrical configuration of the two objects relative to the observer
and by the optical properties of the Venus stratified atmosphere.  
This section addresses 
the visual phenomenon as contemplated from a geocentric vantage point for the 
conditions of the June 2012 transit. 
The geometrical parameters needed in the simulations are either adopted or 
inferred from the JPL HORIZONS service (Giorgini et al., \cite{giorginietal1996})
or the New Horizons GeoViz software.\footnote{
H. Throop, http://soc.boulder.swri.edu/nhgv/}

\subsection{\label{geometrical_sec}Some geometrical considerations}

\begin{figure*}
   \centering
   \includegraphics[width=7cm]{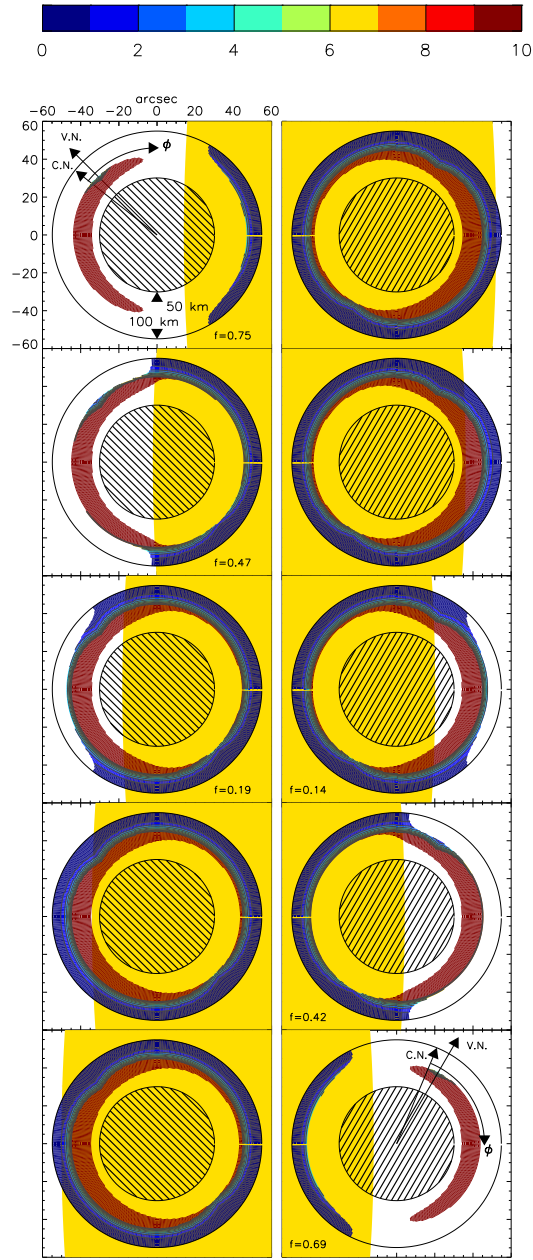}
      \caption{
      Isolines of $\Phi$=$-\log_{10}(L_D\mathbf{(x_O,s_O)}/B_{\odot})$
      at 0.55 $\mu$m during ingress and egress. 
      The Venus
      solid core plus atmosphere up to 50 km are represented to scale in dashed fill.
      The atmospheric ring from 50 to 100 km has been resized for a better
      appreciation of the halo's structure. The atmosphere above 100 km is not shown.
      The unrefracted view of the 
      solar disk is shown to scale in yellow in the background. 
      The directions for the Venus North Pole and the Celestial North are
      indicated. Also indicated is the azimuthal angle $\phi$ as defined in 
      Figs. (\ref{limbcurve_fig})--(\ref{limbcurve_b_fig}). 
      In the color code
      of the figure, blue indicates bright and brown indicates faint.
      For a color version of the figure, see the electronic
      journal.
      }
      \label{halo_seq_shape_fig}
   \end{figure*}

Figure (\ref{halo_seq_shape_fig}) shows a sequence of simulations of both 
the ingress and the egress.  
Each simulation corresponds to a value for the $f$ factor introduced by 
Link (\cite{link1969}) and used by Tanga et al. (\cite{tangaetal2012}). This
factor stands for the fraction of the planet's diameter along the 
Venus-Sun line of centres beyond the Sun's limb. Thus, $f$= 1, 0.5 and 0 are
appropriate for external contact, mid-ingress/-egress and internal contact,
respectively. 
 For a better visualization, 
the atmospheric ring from 50 to 100 km altitude is
stretched by a fixed factor in all images. 
Ignoring for the time being the sunlight extinction caused by the gas and haze,
the Sun's image at the planet's outer terminator appears
notably diminished with respect to the apparent size of the 
unocculted solar disk. 
The diminishment constitutes the visual representation 
of the attenuation by refraction.

Referring to Table (\ref{anglerefr_table}), 
the deflected angle for rays crossing the atmosphere 
with an impact distance $r_b$$-$$R_{\venus}$=55 km is 
$\alpha_{\rm{refr}}$$\sim$1$^{\circ}$,
thus exceeding the solar angular diameter as viewed from Venus
($\sim$2$R_{\odot}$/0.726 AU$\sim$0.734$^{\circ}$).
This means that, in geometrical terms, the halo representations
of Fig. (\ref{halo_seq_shape_fig}) are  the refracted 
images of the entire solar hemisphere visible from Venus.
The innermost/outermost contours 
are the projections of the farther/closer edges of the solar disk,
respectively.

The bottom left and top right images of Fig. (\ref{halo_seq_shape_fig})
show the planet shortly after contact II and before contact III, respectively.
At those instants
the atmosphere does not form a refracted image 
at the inner terminator below $\sim$76 km
on the axis between the centres of the two objects.  
At the outer terminator, a refracted image occurs only above $\sim$59 km. This 
asymmetry
is caused by the unequal distance from each terminator to the opposite
solar limb. The oval ring within which a solar image is not formed dictates the 
range of atmospheric altitudes that cannot be accessed from Earth 
regardless of atmospheric opacity. 
This refraction exclusion ring becomes circular as the planet moves towards mid-transit.

\subsection{Shape and brightness of the halo}

The simulations of Fig. (\ref{halo_seq_shape_fig}) 
assume a continuum wavelength of 0.55 $\mu$m and the
latitude-dependent extinction coefficient described in 
{\S}\ref{venusatmosphere_sec}.
To take into account the varying altitude of clouds along the terminator,  
we solved the RTE over the entire Venus disk for a range of $h_{\rm{cloud}}$ 
values and, subsequently, 
composed the displayed images by assembling 
the relevant latitude--$h_{\rm{cloud}}$ sectors.

Venus enters the transit with its rotational axis tilted by
about 45$^{\circ}$ clockwise with respect to the Venus-Sun line of centres.
At egress, the tilt is about 60$^{\circ}$ and anticlockwise.  
The tilts were estimated graphically from GeoViz simulations.
In combination, the inclination of the rotational
axis and the latitude-dependent altitude of the cloud tops 
break the symmetry of the halo expected from purely geometrical arguments.
The asymmetry is clearly discerned in the contours of 
Fig. (\ref{halo_seq_shape_fig}), each of them corresponding to a value for  $\Phi$=$-\log_{10}(L_D\mathbf{(x_O,s_O)}/B_{\odot})$. 
 
At ingress, the earliest evidence of the halo occurs over the Venus 
North Pole and appears detached from the solar disk until about mid-ingress. 
The bright cap finally turns into a full ring that encircles the 
planet before contact II. 
Essentially, the egress reproduces the ingress sequence in the reverse order. 
During egress, the larger tilt of the rotational axis causes the halo to
appear attached to the solar disk for a longer duration.

Figure (\ref{halo_seq_shape_fig}) provides insight into the 
instantaneous vertical extent of the halo. 
Importantly, the halo's outermost contour 
remains confined to altitudes less than 100 km for most of the ingress/egress. 
The statement is easily verified with a quick estimate specific to 
the instant halfway through ingress/egress. 
At that moment, a  ray emanating from the nearer solar
limb and propagating on the azimuthal plane that includes the Venus-Sun 
line of centres must 
be deflected by an angle 
$\sim$$R_{\venus}$/0.726 AU+$R_{\venus}$/0.274 AU$\approx$42 arc sec to 
reach the geocentric observer. 
Refraction yields such a deflection 
at $\sim$80 km altitude, as seen in Table (\ref{anglerefr_table}).
Thus, at mid-ingress or -egress only rays traced from $\mathbf{x_O}$
on the plane containing the line of centres 
with $r_b$ less than that may connect the observer and the solar disk.

The conclusion serves to highlight the 
importance of haze on the halo's changing appearance 
because the atmosphere is likely optically thick in limb viewing 
at those altitudes. % forming the refracted image of the solar disk.
The halo's inner contour is largely dictated by extinction in
the optically thicker layers of the atmosphere. Overall, the altitudes
contributing effectively to the halo range over 1--2 scale heights.

\subsection{The imaged solar disk}
It is interesting to trace the rays forming the halo 
back to the solar surface elements they departed from.
In geometrical terms, the halo is 
the refracted image of most of the solar disk visible from Venus or Earth.
However, haze extinction concentrates the rays effectively contributing
to the halo's brightness onto a narrow strip near the solar limb. 
Our ray tracing scheme does indeed show that during ingress/egress the halo is 
essentially formed by the solar disk occulted behind Venus. 
That solar region is severely limb-darkened, 
which adds a significant radial gradient to the solar output brightness. 
In the serendipitous situation that a dark sunspot is occulted by the planet during
ingress/egress, one may expect that the sunspot will turn up as a dimmer halo.

\subsection{\label{chromaticity_sec}
Chromaticity of the halo and limb-integrated lightcurves}

The halo's shape and brightness depend on the atmospheric refractivity, 
sunlight extinction and solar limb-darkening. 
In turn, these factors depend on wavelength. 
Focusing on the visible and near-infrared regions of the spectrum,
we investigated them at 0.43, 0.55 and 0.8 $\mu$m, 
which roughly match the center wavelengths of
the B, V and I filters, respectively. 
 
The refractivity of the Venus atmosphere varies with wavelength, which may
introduce a relative displacement between the images formed at different 
colors. 
Considering again Fig. (\ref{sketch_fig}), we can devise two 
 rays of wavelengths $\lambda_1$ and $\lambda_2$
departing from a common solar surface element, 
both of them reaching the observer. 
Since the overall deflections are nearly identical, we can readily estimate
their altitudes of closest approach in the atmosphere. 
According to the approximate, analytical formula given by Baum \& 
Code (\cite{baumcode1953}) for the refraction angle in an isothermal atmosphere,
the number densities at 
closest approach of the two rays are in the ratio 
$n_1$/$n_2$$\sim$$\nu_{\lambda_2}$/$\nu_{\lambda_1}$, with $\nu$ being the
refractivity at the specified wavelength.
For Venus, this means that if 
$\lambda_1$=0.43 and $\lambda_2$=0.8 $\mu$m the rays have their closest 
approach at densities in the ratio $n_1$/$n_2$$\sim$0.9747. 
For an atmospheric scale height of 4 km, this means that the altitudes of closest approach 
differ by about one hundred meters. 
Thus, the displacement of the images due to refraction is relatively minor.

Haze dominates the atmospheric opacity
from the UV to the NIR at the altitudes of formation of the halo.
Recalling from Fig. (\ref{gamma60km_fig}) the spectral dependence 
of $\gamma(\mathbf{x})$, 
it is apparent that the sunlight emerging from the Venus terminator 
must exhibit a moderate red tone imprinted by haze extinction.
The reddening is, however, much less than expected for a Rayleigh atmosphere, 
in which case $\gamma(\mathbf{x})$$\sim$$\lambda^{-4}$. 
Sunlight emitted from the solar limbs is eminently red, as seen in 
Fig. (\ref{U_fig}). 
Thus, solar limb darkening must also contribute towards the reddening of 
the halo. 

%\begin{eqnarray*}
%{{<{L_D}}}({\bf{x_O}}, {\bf{s_O}})/B_{\odot}> =  (
%\int_{\mu_{\rm{M}}}^{\mu_{\rm{m}}}  {{L}}({\bf{x_O}}, {\bf{s_O}})/B_{\odot} d (\mu^2 / 2)
%) /
%(
%\int_{\mu_{\rm{M}}}^{\mu_{\rm{m}}} d (\mu^2 / 2)
%)= 
%\end{eqnarray*}
%\begin{equation}
%=  \int_0^1 L(\xi,\eta)/B_{\odot} d\xi,
%\end{equation}

To study this in a quantitative manner, we produced
radially-averaged radiances of the halo as follows: 
\begin{equation}
\langle \frac{L_D(\bf{x_O},\bf{s_O})}{B_\odot} \rangle = 
\frac{\int\limits_{\mu_M}^{\mu_m} \! L_D(\bf{x_O},\bf{s_O}) / 
\rm{\it{B}_\odot} \, \it{d}(\mu^2/2)}{\int\limits_{\mu_M}^{\mu_m}{d}(\mu^2/2)} 
= \int\limits_{0}^{1} \! {\frac{L_D(\xi,\eta)}{B_\odot}} \, \it{d}\xi
\end{equation}
which is equivalent to averaging over altitude from the ground to 200 km or over
1 arcsec for ground-based observations. Thus, 
$<$${{{L_D}}}({\bf{x_O}}, {\bf{s_O}})/B_{\odot}$$>$ is a valid measure of the
vertically-unresolved brightness with respect to the brightness 
at the Sun's centre. 

\begin{figure*}
   \centering
   \includegraphics{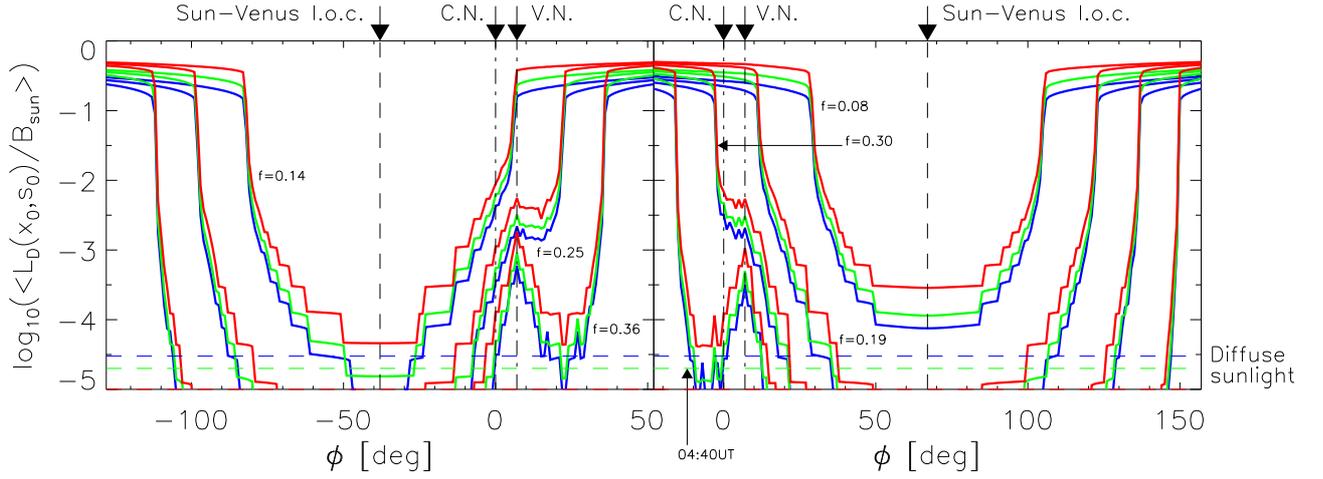}
   \caption{Limb-integrated lightcurves during ingress and egress at 0.43
   (blue), 0.55 (green) and 0.8 (red) $\mu$m. Reference 
   $h_{\rm{cloud}}$--latitude profile. For a color version of the figure, see the electronic
      journal.}
              \label{limbcurve_fig}
% /VENUSTRANSIT/HALO/relflux.pro
    \end{figure*}

Curves of the radially-averaged radiance 
for $\lambda$=0.43, 0.55 and 0.8 $\mu$m during ingress and egress are 
graphed against the azimuthal angle $\phi$ in Fig. (\ref{limbcurve_fig}). 
The curves show a strong time-dependence and reveal with an increased brightness
the lower clouds occurring at the Venus North Pole. 
The color dependence of the lightcurves near the
Sun-Venus line of centres is due to both solar limb darkening and atmospheric 
sunlight extinction. Close to the solar disk edge (the left and right
pedestals of the figure), the higher atmospheric altitudes contribute 
significantly and only solar limb darkening introduces a color dependence.

\begin{figure*}
   \centering
   \includegraphics[width=9cm]{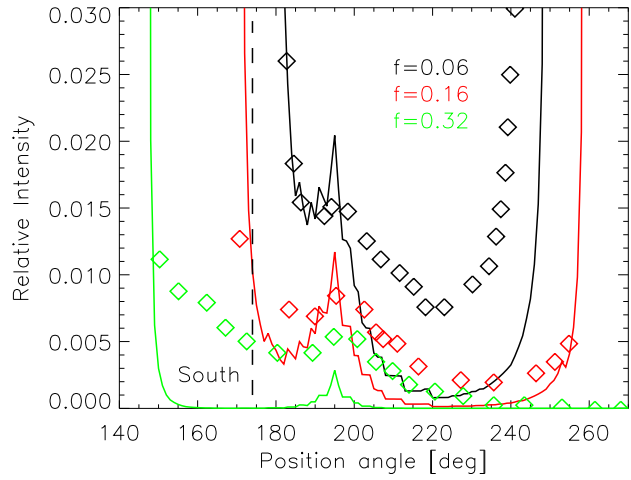}
      \caption{Relative intensity against position angle from G-band observations
      (symbols) at the Dutch Open Telescope on La Palma (Spain)
      and modelling in this work at 0.43 $\mu$m (solid) for the egress during the 2004 Venus
      transit. The intensities and position angle are defined
      by Tanga et al. (\cite{tangaetal2012}). To reproduce the peak in
      brightness away from the South Pole, we needed to shift the 
      reference $h_{\rm{cloud}}$--latitude profile by about 20$^{\circ}$.
      For a color version of the figure, see the electronic journal.
      }
         \label{limbcurve2004_fig}
% /VENUSTRANSIT/HALO/halo_map2004.pro
% /VENUSTRANSIT/HALO/relflux2004.pro
   \end{figure*}

These curves constitute a direct target for comparison with 
photometric measurements. Indeed,
Tanga et al. (\cite{tangaetal2012}) recently published limb-integrated curves 
from observations of the 2004 transit. 
Figure (\ref{limbcurve2004_fig}) shows with symbols 
their G-band lightcurves obtained during egress 
at the Dutch Open Telescope on La Palma as digitized from their 
Fig. 8. Relative intensities are represented against the azimuthal angle 
and reveal the expected U shape as well as a peak in brightness displaced from the pole. 
The solid curves are our model lightcurves at 0.43 $\mu$m for the 2004 egress
conditions. 
Tanga et al. (2012) normalize
their lightcurves with the solar disk brightness at about one Venus radius from
the solar edge. 
For consistency, we re-scaled our radially-averaged radiances by
the appropriate limb-darkening factor, so that the two sets of curves become
directly comparable. The comparison for three different instants during egress 
is overall satisfactory, especially taking into account that we assumed a 
global description for the cloud top altitude. 
Importantly, to achieve such an agreement, we needed to offset 
the location of the minimum cloud top altitude by about 20$^{\circ}$ away 
from the pole.
If confirmed, that feature might indicate that occasionally 
the cloud top altitude does not descrease monotonically with latitude towards
the poles. 

Following the comparison with Tanga et al. (2012),
that work reports that as factor $f$ increases the halo brightness decays more
rapidly at red wavelengths than at visible and ultraviolet wavelengths. 
As discussed earlier, our model predicts an overall red halo, 
which is somewhat at odds with the Tanga et al. (2012) findings. 
At any rate, the amplitude of the error bars in 
their experiment might also accomodate a small reddening. New observations of
the 2012 transit will hopefully provide a better insight into the issue.

\begin{figure*}
   \centering
   \includegraphics{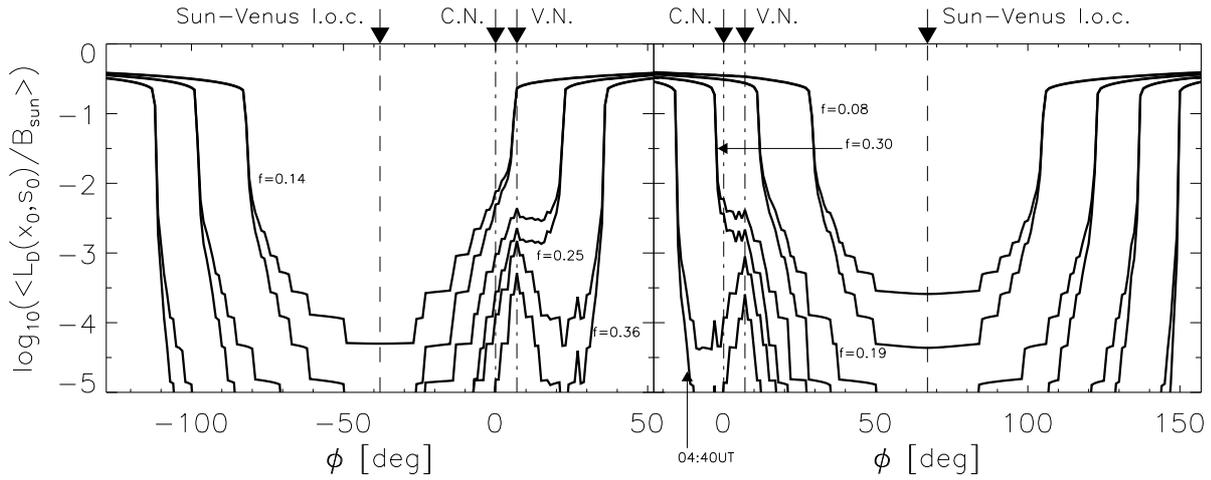}
   \caption{Limb-integrated lightcurves at 0.55 $\mu$.
   Each pair of curves corresponds to the lightcurves obtained from perturbing the reference 
   $h_{\rm{cloud}}$--latitude profile by $\pm$1 km.
   }
              \label{limbcurve_b_fig}
% /VENUSTRANSIT/HALO/relflux.pro
    \end{figure*}

One may consider to use model lightcurves such as those in Fig. 
(\ref{limbcurve_fig}) to 
fit the empirical lightcurves and retrieve cloud altitudes along the planet's
terminator. 
Tanga et al. (\cite{tangaetal2012}) have already approached the task. 
To understand to what extent the model
lightcurves depend on the prescribed atmospheric optical properties, 
we produced limb-integrated lightcurves 
with the cloud top altitude shifted at all latitudes 
by $\pm$1 km with respect to the reference cloud altimetry. 
The resulting lightcurves are given in Fig. (\ref{limbcurve_b_fig}) and show
that shifting the cloud tops by $\pm$1 km leads to variations in
the limb-averaged radiances by factors of $\sim$2 or more. 
Ultimately, the precision of the method to retrieve the cloud top altitude 
will depend strongly on the quality of the photometric data. 
Spatial resolution in the data is not anticipated to be critical, since 
resolutions of about 1 arcsec are commonly achieved also from ground-based 
observatories. A cadence of one image per minute
or less suffices to reveal the temporal evolution of the lightcurves. 

Some observers of the 2012 Venus transit may also attempt the spectroscopic 
characterization of the halo. The likely result of such an experiment 
will be similar to a transmission spectrum of the Venus atmosphere from an
altitude of 85 km or higher with a continuum dependent on wavelength through
both haze extinction and solar limb darkening.

%Hedelt et al. (\cite{hedeltetal2011}) made spectroscopic measurements of
%the planet's limb within the transit. One may devise that the same observing
%approach could be applied to the ingress/egress phases of the transit. 

%The SOIR instrument on Venus Express (Vandaele et al., \cite{vandaeleetal2008})
%is producing limb-viewing transmission spectra of the Venus atmosphere from
%solar occultations. So far, the SOIR instrument has detected molecules such as 
%CO$_2$ (a few isotopologues), H$_2$O, HDO, HF, HCl, ... (Ref., ...) 
%and set upper limits to other molecules such as XXX, (Ref., ...). 
%High spectral resolution observations of the halo may provide suitable
%conditions for further exploration of some of those targets or even for the
%detection of the O$_2$ molecule (Ref., ...). \textbf{what's the Doppler shift of
%the planet??}

\subsection{Diffuse sunlight contribution to the halo} 

One more issue to explore is the contribution of diffuse 
sunlight to the overall brightness of the halo. 
In between external and internal contacts, 
the Sun-Venus-Earth angle is more than $\sim$179$^{\circ}$, which favors forward
scattering from the planet's terminator.  
%Thus, particles in suspension in the atmosphere
%scatter the incident sunlight in a nearly forward scattering configuration. 
Most atmospheric aerosols scatter efficiently in the forward direction, with the 
scattering efficiency depending strongly on the ratio 
$x_{\rm{eff}}$=2$\pi$$r_{\rm{eff}}$/$\lambda$ of the effective radius for the 
particle size distribution and the incident sunlight wavelength (Hansen \& Travis, 
\cite{hansentravis1974}). As seen in Fig. (\ref{phF_fig}), 
the peak in the scattering phase function 
becomes enhanced at the shorter wavelengths and is particularly stronger 
for the larger mode-2 haze.  
 
\begin{figure*}
   \centering
   \includegraphics[width=9cm]{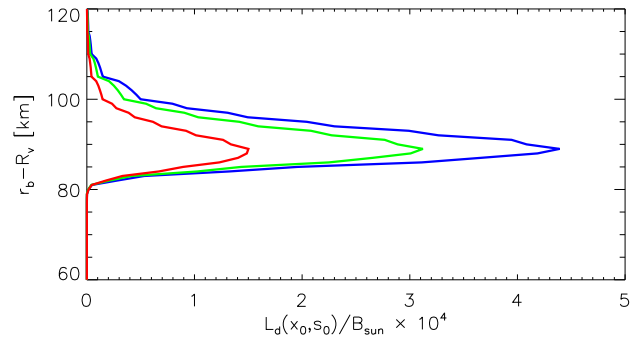}
      \caption{Diffuse radiance $L_d\mathbf{(x_O,s_O)}/B_{\odot}$ graphed 
      against the impact radius $r_b - R_{\venus}$. From right to left, the
      curves correspond to wavelengths of 0.43, 0.55 and 0.8 $\mu$,
      respectively. For a color version of the figure, see the electronic
      journal.}
      \label{diffuse_fig}
% /VENUSTRANSIT/DIFFUSE
   \end{figure*}

Figure (\ref{diffuse_fig}) graphs the multiple-scattering diffuse 
radiance at mid-transit (with the Sun, Venus and Earth perfectly aligned) 
against the impact radius as calculated with the Monte Carlo method of 
{\S}\ref{diffuse_sec}. For simplicity, we assumed that the cloud top lies 
at a constant altitude of 70 km at all latitudes. 
The radiances peak at $\sim$88 km, 
that corresponds approximately to the $\tau_{\rm{limb}}$=1 level. 
Shifting upwards/downwards the clouds by a few kilometers 
merely shifts the radiances by the same distance but has a negligible impact on
the vertically-integrated magnitudes.

Integration of the curves in Fig. (\ref{diffuse_fig}) results in 
radially-averaged diffuse radiances $<$$L_d\mathbf{(x_O,s_O)}/B_{\odot}$$>$ of 
$\sim$3$\times$10$^{-5}$, 2$\times$10$^{-5}$ and 1$\times$10$^{-5}$ at 0.43, 0.55 and
0.8 $\mu$m, respectively. Comparison with the lightcurves of 
Fig. (\ref{limbcurve_fig}) indicates that refracted sunlight dominates the halo's
brightness over most of the ingress/egress.

\section{\label{transit_sec} Venus-Sun disk-integrated lightcurves} 

We have thus far assumed that the Venus disk is spatially resolvable 
from Earth, a necessary condition to investigate the 
connection between the halo's structure and the cloud altimetry along the
planet's terminator.
However, the question of whether the
atmospheric properties of a transiting planet can be disentangled 
from the planet-star blended signal has become key in the research of 
extrasolar planets (Seager \& Sasselov, 
\cite{seagersasselov2000}; Brown, \cite{brown2001}; Hubbard et al., 
\cite{hubbardetal2001}). 

The so-called technique of transits aims to identify the signatures of the
planet and its atmosphere from comparative measurements of the planet-star 
light out of and in transit.
In-transit spectroscopy (or spectrophotometry) is providing 
valuable insight into 
exoplanetary atmospheres, especially for the giant planets that
constitute most of the exoplanet detections to date. 
As the sizes of discovered extrasolar planets shrink and technology
improves, the prospects for 
characterizing Earth-sized planets with this technique, 
including Venus-like ones, 
look closer. % and closer. 
Next, we describe the Venus transit as if observed in integrated 
light from a terrestrial distance and from a remote vantage point. 
 
\subsection{\label{diskintlightcurve_sec}Disk-integrated lightcurves from Earth and far away} 
 
\begin{figure*}
   \centering
   \includegraphics[width=9cm]{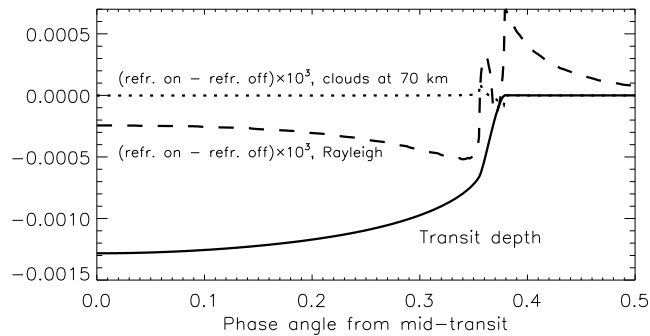}
      \caption{Transit depths for the Venus-Sun integrated signal as observed
      from Earth. The dashed curves represent the differential transit depth
      contributed by refraction.}
      \label{lc1AU_fig}
% /VENUSTRANSIT/DIFFUSE
   \end{figure*}

Figure (\ref{lc1AU_fig}) shows model transit depths  at 0.55 $\mu$m 
for the Venus-Sun system as observed from Earth. 
The transit depths are expressed in units of irradiance for the
unobstructed Sun. The overall shape of the 
lightcurve is determined by the ratio of Venus/Sun solid angles,
$\sim$1/30, and the limb darkening function at the specified wavelength.

Refraction introduces unique features in the lightcurves near external and
internal contacts. In Fig. (\ref{lc1AU_fig}), we show also the difference in
transit depths when refraction is taken into account and when it is omitted. 
If the Venus atmosphere was free of haze and cloud particles, 
the differential transit depth curve would show 
that the planet's size is determined by a combination of Rayleigh scattering and
refraction. 
A refraction exclusion ring 
appears at altitudes from $\sim$59 to $\sim$76 km, depending on the
distance from the planet's local terminator to the solar limb, 
and contributes most significantly to the disk-integrated lightcurve near
internal contact. Note also that the depth varies by
$\sim$3$\times$10$^{-7}$ from mid-transit to internal contact, which means 
that the planet's measurable radius apparently expands by a few km. 
Out of transit, the transit depth inverts its sign because 
sunlight refracted by the planet contributes positively to the brightness of the
Venus-Sun system. The magnitude of this additional brightening amounts to 
$\sim$7$\times$10$^{-7}$ at external contact. 
Including haze in the calculation of the optical opacity reduces by orders of
magnitude the effects of refraction in the differential transit depths, as the
dotted curve in the figure indicates. 
At any rate, such small perturbations fall well below 
typical irradiance fluctuations due to solar
oscillations over periods of minutes (Kopp et al., \cite{koppetal2005}).

\begin{figure*}
   \centering
   \includegraphics[width=9cm]{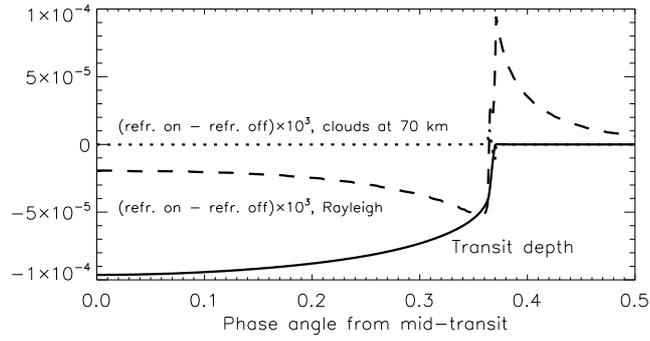}
      \caption{Same as Fig. (\ref{lc1AU_fig}) with the observer at an infinite
      distance.}
      \label{lcINF_fig}
% /VENUSTRANSIT/DIFFUSE
   \end{figure*}

The disk-integrated lightcurves measured by a remote observer would exhibit the
same phenomena described above. The considerably smaller ratio of Venus/Sun 
solid angles in that case, $\sim$$R_{\venus}/R_{\odot}$$\sim$1/115, would cause
the same phenomena to appear about one order of 
magnitude fainter in the combined planet-star signal. 
Figure (\ref{lcINF_fig}) demonstrates the disk-integrated lightcurves in the limit of an
infinitely distant observer.

\subsection{Disk-integrated transmission spectra}

The monochromatic lightcurves presented in {\S}\ref{diskintlightcurve_sec} 
might also be calculated at wavelengths exhibiting discrete molecular 
absorption features. 
We produced spectra for the transit depth at wavelengths from about 
0.3 to 5 $\mu$m. In addition to CO$_2$, the atmosphere was assumed to contain
small amounts of N$_2$, CO and O (Vandaele et al., \cite{vandaeleetal2008};
Hedin et al., \cite{hedinetal1983}) and H$_2$O (1 ppm; Fedorova et al.,
\cite{fedorovaetal2008}). 
The line parameters for molecular transitions were borrowed from HITRAN
2008 (Rothman et al., \cite{rothmanetal2009}).  
The transit depths were degraded to
a resolving power of about 10$^3$ and expressed in terms of the
equivalent height. (The equivalent height is the height of an opaque slab that
would produce an identical transit depth.) 
In equivalent heights, the spectra measured from Earth and from a
remote distance are nearly identical, yet the associated stellar dimming 
may differ by an order of magnitude. Whether observed from Earth
($d_{\rm{\bf{O}}}$/$l_{\odot}$$\sim$0.274) or from a remote distance
($d_{\rm{\bf{O}}}$/$l_{\odot}$$\sim$1), the solar irradiance blocked by 
the Venus disk and atmosphere is 
$F_{d+a}$$\sim$$\pi((R_{\venus}+h_{\rm{eq}})/d_{\rm{\bf{O}}})^2$$B_{\odot}$$U(\mu_{\odot})$,
where we tacitly assume that the planet is nearly point-like and the background
solar radiance is determined by a single $\mu_{\odot}$ dependent on the planet's
phase. In relative terms, the solar dimming during transit is  
$F_{d+a}$/$F_{\odot}$.

%$((R_{\venus}+h_{\rm{eq}})/R_{\odot})^2$$\times$$(l_{\odot}/d_{\rm{\bf{O}}})^2$$\times$$U(\mu_{\odot})$/$(\sum_{i=0}^5 2u_i/(i+2))$. 

\begin{figure*}
   \centering
   \includegraphics[width=9cm]{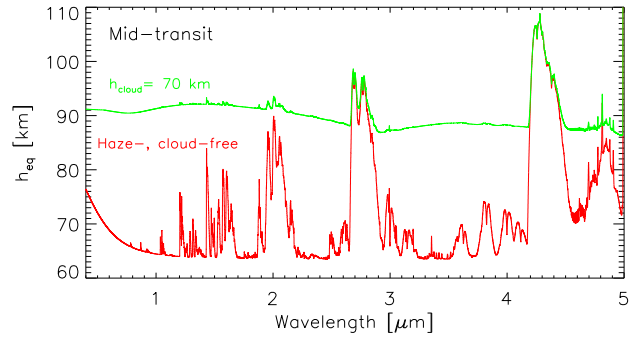}
      \caption{Mid-transit transmission spectrum of Venus at a resolving power 
      of 10$^3$ in two different atmospheric conditions. 
      For a color version of the figure, see the electronic journal.
}
      \label{spectrum_fig}
% /VENUSTRANSIT/DIFFUSE
   \end{figure*}

The curves in Fig. (\ref{spectrum_fig}) correspond to mid-transit ($\mu_{\odot}$=1) spectra calculated 
for an atmosphere free of haze and cloud particles 
and for a hazy atmosphere with cloud tops at 70 km. 
The latter seems to reproduce most of the structure of the in-transit
spectrum presented by Ehrenreich et al. (\cite{ehrenreichetal2012}). 
The most obvious difference between the Rayleigh and hazy atmospheres 
refers to the continuum level, that lies considerably higher in the latter. 
For the Rayleigh atmosphere, the continuum level is largely 
determined by refraction at wavelengths longer than $\sim$1 $\mu$m, 
with Rayleigh scattering contributing shortwards of that. 
For the hazy atmosphere, the continuum is dictated by the haze 
optical properties throughout the spectrum. 
The simulations for phases other than mid-transit (not shown) 
indicate that for the Rayleigh atmosphere the transit depth of molecular bands 
is highest at mid-transit and a few kilometers less when the planet is near contacts. 
This modulation is attributable to refraction, and becomes entirely negligible for the hazy atmosphere.

The comparison is relevant because to some extent a clear-atmosphere Venus 
would be akin to our own Earth. 
Regarding the molecular structure of the spectra, it is overwhelmingly 
dominated by rovibrational bands of CO$_2$. 
Benneke \& Seager (\cite{bennekeseager2012}) have 
investigated what can unambiguously be inferred from in-transit spectroscopy of
a super-Earth, and noted that the lacking evidence of
Rayleigh scattering may impede the identification of the main atmospheric
constituent. 
Interestingly, it appears that the in-transit spectrum of Venus
would reveal little on the identity of the main atmospheric constituent.

\begin{figure*}
   \centering
   \includegraphics{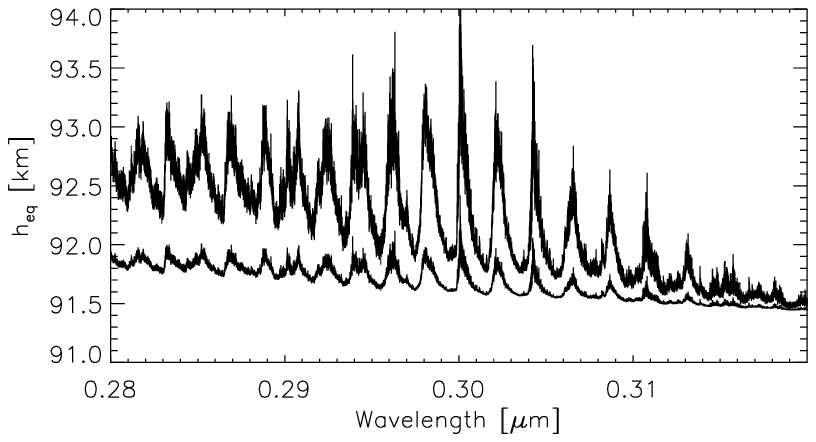}
   \caption{Equivalent height for altitude-independent SO$_2$
   mixing ratios of 1$\times$ and 5$\times$10$^{-7}$. 
   The SO$_2$ absorption cross sections are from Blackie et al.
   (\cite{blackietal2011, blackietal2011b})}
              \label{spectrumSO2_fig}
% /VENUSTRANSIT/HALO/relflux.pro
    \end{figure*}

A final comment on the SO$_2$ molecule. 
The SO$_2$ molecule plays a key role in the 
Venus photochemistry and in the formation of H$_2$SO$_4$ in the clouds
(Mills \& Allen, \cite{millsallen2007}). 
The SO$_2$ molecule absorbs at wavelengths below $\sim$0.32 $\mu$m, at the limit of accessibility
for ground-based observations imposed by the terrestrial ozone cut-off at $\sim$0.3 $\mu$m. 
Belyaev et al. (\cite{belyaevetal2012}) have retrieved the SO$_2$ mixing ratio for altitudes below 105 km.
The precise amounts appear to be sensitive to the retrieval approach, which leads to values 
somewhere between 10$^{-7}$ and 10$^{-6}$ at altitudes above the $\tau_{\rm{limb}}$=1 level of $\sim$90 km. 
Figure (\ref{spectrumSO2_fig}) shows the equivalent height near 0.3 $\mu$m at high resolving power for
SO$_2$ mixing ratios of 10$^{-7}$ and 5$\times$10$^{-7}$. The main conclusion to obtain from the graph is that SO$_2$
introduces a distinct signature that might be captured in spectroscopic observations of the transit if enough signal-to-noise
ratio is achieved.

\section{\label{conclusions_sec}Summary} 
 
The current work presents theoretical insight into the signature imprinted 
on sunlight by the Venus atmosphere during the passage of the planet 
in front of the solar disk. 
The conclusions are specific to the June 2012 event but are easily 
generalizable to other transits. Our model approach includes 
refraction, multiple scattering and extinction by gases and aerosols 
in the atmosphere, as well as solar limb darkening, 
and handles all phases of the transit 
(including the out-of-transit, ingress/egress and in-transit stages) 
in a consistent manner. 
To the best of our knowledge this is the first theoretical
investigation of the Venus transit at this level of detail.
 
We investigated the halo that forms at the planet's outer terminator near 
ingress/egress, predicting its shape, brightness and chromaticity. 
Both the inner and outer contours of the halo are considerably affected by 
haze in the atmosphere above the clouds and below 100 km. 
The impact of haze opacity on the halo structure explains the 
variability of halo patterns from ingress to egress and from event to event. 
We estimated that scattered sunlight contributes less than refracted sunlight
for most of the ingress/egress between internal and external contacts. 
Since our predictions are based on a realistic description of the 
Venus atmosphere it should be possible to verify them provided that accurate
measurements become available. 
Further, our formulation
opens a way for the characterization of the Venus atmosphere from observations
during the transit. 
In retrospective, it is fair to state that if the Venus haze extended 
for 2--3 scale heights higher than it does, 
it is likely that the identification of a Venus
atmosphere in the 18th century would have to have waited longer.

We presented lightcurves for the combined Venus-Sun signal during transit and 
discussed some features of the lightcurves associated with refraction. 
We also presented model in-transit spectra of Venus and 
of a haze- and cloud-free Venus-like planet. 
Haze appears recurrently as a key element in the interaction of
sunlight and the Venus atmosphere. 
The disparate vertical extent of aerosols in their atmospheres
establishes a considerable difference between Venus and Earth.
Had the proper technology existed, it might have been appropriate 
to investigate the historic record of observed Venus transits. 
In this respect, the 2004 and 2012 events may set a valuable start point for 
comparisons between transits.
% we could explore the effect of tilt in past events...

Finally, the Venus transit of 2012 arrives at a moment of great activity in 
the exploration of extrasolar planets by the method of transits. We have seen
that high-altitude aerosols introduce specific challenges to the
characterization of the atmospheric composition and cloud top altitude for 
a transiting Venus. 
Since it is still unknown how frequently high-altitude clouds may occur in the
atmospheres of extrasolar planets, 
there are reasons to think that the challenges posed by Venus might also be 
generally encountered in the characterization of those planets. 
Indeed, recent work on hot Jupiter HD 189733b (Lecavelier des Etangs et al.,
\cite{lecavelierdesetangsetal2008}; Huitson et al., \cite{huitsonetal2012}) 
and super-Earth GJ 1214b 
(Berta et al., \cite{bertaetal2012}) suggest that haze must be invoked to 
explain the appearance of their transmission spectra.

%\cleardoublepage
\newpage

\begin{table*} 
\caption{\label{anglerefr_table} Total refracted angles $\alpha_{\rm{refr}}$ for
rays at 0.55 $\mu$m entering the atmosphere with an impact radius $r_b$. For
comparison, $\alpha_{\rm{refr}}^*$ is the refracted angle for an isothermal
atmosphere of  refractivity $\nu_{\lambda=0.55\mu m}$ and 
scale height $H_{\rm{atm}}$ (Baum \& Code, \cite{baumcode1953}). 
}
\begin{flushleft}
\begin{tabular}{ccccc}
\hline
$r_b-R_{\venus}$ & $\nu_{\lambda=0.55\mu m}$ & $H_{\rm{atm}}$  & $\alpha_{\rm{refr}}$ 
& $\alpha_{\rm{refr}}^*$ \\
\multicolumn{1}{c}{[km]} & \multicolumn{1}{c}{} & [km] &\multicolumn{2}{c}{[arc sec]} \\
\hline
\hline
55  & 2.12$\times$10$^{-4}$ & 6.6 & 3794 & 3444  \\
60  & 1.09$\times$10$^{-4}$ & 5.6 & 2048 & 1852  \\
70  & 1.82$\times$10$^{-5}$ & 5.0 & 325 & 328  \\
80  & 2.56$\times$10$^{-6}$ & 4.5 & 45 & 48  \\
90  & 2.65$\times$10$^{-7}$ & 3.8 & 5.1 & 5.5  \\
100 & 1.91$\times$10$^{-8}$ & 3.8 & 0.45 & 0.39  \\
110 & 1.47$\times$10$^{-9}$ & 4.1 & 0.036 & 0.029  \\
120 & 1.26$\times$10$^{-10}$ & 4.3  & 0.0030 & 0.0024  \\
\hline
\multicolumn{5}{l}{
$\alpha_{\rm{refr}}^*$= 
$\nu_{\lambda=0.55\mu m}$$\sqrt{2\pi R_{\venus}/H_{\rm{atm}}}$.  
} \\
%\tableline
%\tableline
\hline
\end{tabular}
\end{flushleft}
\end{table*}

%
%______________________________________________________________

\begin{acknowledgements}
      AGM gratefully thanks Thomas Widemann for the invitation to the 
      3rd Europlanet strategic workshop -- 4th PHC/Sakura meeting: Venus as a
      transiting exoplanet, held in Paris on 5--7 March 2012, and  
      Agust\'in S\'anchez-Lavega and the Grupo de Ciencias Planetarias at
      the UPV/EHU for hospitality during the elaboration of the work.
      Finally, the authors acknowledge the referee's authoritative
      report, which helped improve the manuscript. 

\end{acknowledgements}

\end{document}